# Formation of Long Single Quantum Dots in High Quality InSb Nanowires Grown by Molecular Beam Epitaxy

Dingxun Fan,[1] Sen Li,[1] N. Kang,[1,*] Philippe Caroff,[2,†] L. B. Wang,[1] Y. Q. Huang,[1]

M. T. Deng,[3] C. L. Yu,[3] and H. Q. Xu[1,3,*]

[1] *Department of Electronics and Key Laboratory for the Physics and Chemistry of Nanodevices, Peking University, Beijing 100871, China*

[2] *I.E.M.N., UMR CNRS 8520, Avenue Poincaré, BP 60069, F-59652 Villeneuve d'Ascq, France*

[3] *Division of Solid State Physics, Lund University, Box 118, S-221 00 Lund, Sweden*

**Abstract**

We report on realization and transport spectroscopy study of single quantum dots (QDs) made from InSb nanowires grown by molecular beam epitaxy (MBE). The nanowires employed are 50-80 nm in diameter and the QDs are defined in the nanowires between the source and drain contacts on a Si/SiO$_2$ substrate. We show that highly tunable QD devices can be realized with the MBE-grown InSb nanowires and the gate-to-dot capacitance extracted in the many-electron regimes is scaled linearly with the longitudinal dot size, demonstrating that the devices are of single InSb nanowire QDs even with a longitudinal size of ~700 nm. In the few-electron regime, the quantum levels in the QDs are resolved and the Landé g-factors extracted for the quantum levels from the magnetotransport measurements are found to be strongly level-dependent and fluctuated in a range of 18-48. A spin-orbit coupling strength is extracted from the magnetic field evolutions of a ground state and its neighboring excited state in an InSb nanowire QD and is on the order of ~300 µeV. Our results establish that the MBE-grown InSb nanowires are of high crystal quality and are promising for the use in constructing novel quantum devices, such as entangled spin qubits, one-dimensional Wigner crystals and topological quantum computing devices.

Over the past decade, transport measurements of semiconductor quantum dots (QDs) have been widely used for exploring novel physics and new applications. Many-body phenomena, such as the Kondo effect[1-5] and solid-state spin qubits,[6,7] have been extensively studied using semiconductor QDs. These abundant phenomena rely on the coherent nature of electron transport in the systems. More recently, InSb nanowires (NWs) have attracted an increasing interest.[8-18] Owing to the intrinsic properties of bulk InSb, such as a small bandgap $E_g = 0.17$ eV, a high electron mobility $\mu_e = 77000 \, \text{cm}^2/\text{Vs}$, a small electron effective mass $m_e^* = 0.015 \, m_e$ (where $m_e$ is the bare electron mass), and a large electron Landé g factor $|g^*| = 51$,[19,20] InSb nanowires have potential applications in the fields of quantum computation, spintronics and high-speed electronics. Using state-of-the-art nanofabrication techniques, various InSb NW devices, such as field-effect transistors,[10,13,17] single and double QDs,[8-10,12,16] and semiconductor-superconductor hybrid quantum devices,[11] have been realized. Studies of these devices have led to the observations of phase-coherent universal conductance fluctuations,[13] large and energy level-dependent g-factors,[8,9,14] strong spin-orbit interaction strengths,[8,12] and correlation-induced conductance suppression,[9] and to the demonstration of electric manipulation of electron and hole spin states.[12,15] Very recently, Majorana bound states in solid state systems have attracted great attention, because of their potential applications in topological quantum computing, and InSb semiconductor NW-superconductor hybrid quantum devices have been developed to spot the signatures of these exotic, topologically distinctive states.[21-24] Nevertheless, in the aforementioned works, the InSb NWs used have primarily been grown by metal-organic vapor phase epitaxy (MOVPE) or chemical vapor deposition (CVD). Due to their intrinsically high material purity and crystal quality, InSb nanowires grown by molecular beam epitaxy (MBE) are desired for the development of well-defined quantum and topological devices to detect novel, exotic physics phenomena. However, electrical properties of MBE-grown InSb NWs have only been studied at room temperature.[25] Quantum devices made from these presumably high-quality MBE-grown InSb NWs and many-body coherent transport in such devices have yet to be demonstrated.

In this communication, we report on the realization and low-temperature transport measurements of single QDs in MBE-grown InSb NWs. The measurements of the QD devices in the many-electron regime demonstrate regular, consecutive Coulomb diamonds of similar sizes in the charge stability diagrams, indicating that the transport occurs dominantly as single electron tunneling through the QDs. The gate capacitance to the QDs deduced from Coulomb oscillations is found to scale linearly with the spacing between the source and drain contacts, demonstrating that the devices are of single QDs and each is built from the entire nanowire segment between the contacts. In the few-electron regime, the devices are studied by magnetotransport spectroscopy measurements. The g factors are extracted for different quantum levels in the QDs using magnetic-field evolutions of both the differential conductance at different back-gate voltages and the linear-response conductance at zero bias voltage. The measured g factors show level-dependent fluctuations in a range of 18-48, indicating the presence of spin-orbit interaction in the InSb NW QDs. A spin-orbit coupling strength of ~ 275 μeV is extracted from the magnetic-field evolutions of a ground state and its neighboring excited state of an InSb NW QD in this work.

All InSb nanowires used in this study are grown by gas-source MBE using gold seed particles decorating an InP(111)B substrate (Figure 1a). The gold seeds are obtained by dewetting of a 0.3 nm nominally thick gold film at 510 °C. Growth is initiated by growing homoepitaxial InP wurtzite stem nanowire segments for 15 min, followed by growing wurtzite InAs stem segments for 25 min, before switching to grow InSb nanowires for 25 min. Cooling down is performed with a linearly decreasing $Sb_2$ molecular flux during 45s and then in vacuum until the sample is removed from the growth chamber at room temperature. The Sb$_2$ flux is obtained from a Veeco valved-cracker cell. The growth is performed at 410 °C for all segments. The In flux 2D equivalent growth rate is calibrated to 0.50 ML/s on an InAs(001) buffer layer and the total V/III ratio (estimated by the ratio of 2D equivalent growth rates for both group III and group V adatoms) is set to 1.95 for the InSb segment growth. A perfect twin-free zincblende crystal structure is obtained independently of diameter for all InSb nanowire segments. Crystal phase perfection over a large range of growth conditions, nanowire dimensions, or growth techniques, is a unique advantage of the gold-seeded antimonide nanowire family, which has been confirmed independently in several works.[26] Our InSb nanowire segments are also completely free of tapering. See previous works focused on the growth and structural properties for more details on these aspects.[9,25,27]

The as-grown InSb NWs are mechanically transferred onto an n$^{++}$ doped Si substrate covered by 105 nm thick thermally grown SiO$_2$, which serve as a global back gate and gate dielectric, respectively. Standard electron-beam lithography is used to define source and drain electrodes and to connect the electrodes and thus nanowires to pre-defined outer bonding pads. After development, the sample is oxygen-plasma ashed for 15 s to remove the resist residues and then chemically etched in a diluted (NH$_4$)$_2$S$_x$ solution at 35 °C for 1 min to remove the native surface oxide.[28] Then the sample is immediately loaded into the vacuum chamber, followed by thermal evaporation of Ti/Au (5/100 nm) as contact metal and lift-off in acetone and isopropanol (IPA). A quantum dot is formed in an InSb NW segment between the contacts by naturally formed Schottky barriers at cryogenic temperatures.

Figure 1b shows a tilted, false-colored SEM image of a fabricated InSb NW QD device. The diameters of NWs used in our device fabrication are in a range of 50-80 nm. The fabricated devices are characterized at room temperature in a probe station. Devices with two-terminal resistances of 20-100 kΩ are selected and cooled down in a $^3$He/$^4$He dilution refrigerator for low-temperature transport spectroscopy studies. The data are recorded using both dc and ac measurement methods in different cooling-down cycles. In the dc measurements, the voltage is applied antisymmetrically to the source and drain electrodes to suppress common-mode noise. The current is amplified using a home-made current amplifier and is numerically differentiated to obtained the conductance. While in the ac measurements, a 17.3 Hz, 5 μV root-mean-squared excitation voltage source is fed into one of the two contacts of a device and the current is recorded from the other side. The signal lines are carefully filtered and all the measurements are carried out at a base temperature of 60 mK. In the magnetotransport measurements, the magnetic field $B$ is applied perpendicularly to the substrate and thus to the NWs in the devices.

Figure 2a shows the two-terminal current $I_{ds}$ of a device with ~68 nm in the NW diameter and ~391 nm in the contact spacing measured as a function of the back gate voltage $V_{bg}$ at a fixed bias voltage of $V_{ds}$ = 100 μV. It is seen that the measured current $I_{ds}$ shows regular, nearly periodic current oscillations with a series of sharp peaks separated by zero conductance regions, i.e., the single-electron Coulomb blockade (CB) effect in the many-electron regime. We estimate the gate capacitance $C_g$ from the averaged peak-to-peak separation, which is solely determined by the capacitive coupling of the QD to the gate, and find $C_g \approx 16$ aF. Figure 2b shows the differential conductance $dI_{ds}/dV_{ds}$ of the device measured as a function of $V_{ds}$ and $V_{bg}$ (charge stability diagram). Here, the differential conductance is represented in a color scale with bright (dark) color corresponding to high (low) conductance. Figure 2b consists of 14 consecutive, diamond-shaped, dark areas in which the charge transport is prohibited due to the single-electron CB effect. The sizes of the diamonds do not show clear electron-number dependence, indicating negligible contribution of the quantization energy $\Delta\varepsilon$ to the single-electron addition energy $\Delta E$ to the QD, where in the constant interaction model $\Delta E = \Delta\varepsilon + E_C$ and $E_C$ is the single-electron electrostatic charging energy. Hence, we extract from the vertical sizes of the diamonds an averaged charging energy $E_C \sim \Delta E \sim 3.5$ meV, which gives an averaged value for the total QD capacitance $C_\Sigma$ of 45 aF and an averaged value for the gate lever arm factor $\alpha = C_g/C_\Sigma$ of 0.36. Indeed, using an ideal one-dimensional infinitely deep quantum well model, the longitudinal single-particle quantization energy in the nanowire with a length of $L = 391$ nm is $\Delta\varepsilon \sim 0.4$ meV, an order of magnitude smaller than the charging energy. In addition, one can also observe that the sizes of the CB diamonds are slightly smaller in the higher $V_{bg}$ region, possibly due to an increase in the QD size at more positive $V_{bg}$.

The regular, consecutive CB diamond structures seen in Fig. 2b suggest the formation of a single quantum dot in the device. Existence of more dots in series in the device would be expected to show a more complicated overlapping diamond structure in the charge stability diagram.[29-31] This result of the formation of a single quantum dot in a 391-nm-long MBE-grown InSb NW is very encouraging. To support the result, the gate-to-dot capacitances $C_g$ of four other fabricated InSb NW devices are measured. These devices are made from the InSb nanowires of similar thickness (70-80 nm in diameter) but have different source-to-drain contact spacings and the measurements are all performed in the well-defined many-electron regime in which 10 or more consecutive current peaks with an approximately equal spacing are observed in each device. Figure 2c shows the gate-to-dot capacitance $C_g$ extracted from the measurements as a function of the source-to-drain spacing $L$. It is seen that the measured data points are fitted reasonably well to a straight line with a slope of 39±2 aF/μm, indicating the formation of a single QD in each of these devices, including the one with the source-to-drain spacing of ~700 nm. Note that we have also estimated the gate capacitance $C_g$ to an InSb nanowire segment using a cylinder-on-plane model[32] and found values of $C_g$ which are only 20%-30% higher than the corresponding experimental data. Here, the deviations could be understood as the combined effect of screening from the source and drain electrodes and of neglecting the contribution from small but finite quantum level spacings. We would like to emphasize again that these data demonstrate that the QD size is solely determined by the spacing between the source and drain contacts, underscoring the high crystal quality of the MBE-grown InSb nanowires. We also note that the effects of lattice defects, crystallographic imperfections, or impurities in the nanowires, which could break a single QD into a multiple one, have not been observed in these measurements.

Having demonstrated the electron transport though single QDs in the MBE-grown InSb NW devices in the many-electron regime, we now move on to the investigation of the transport properties of the devices in the few-electron regime. Figure 3a presents the measured stability diagram of a QD device made with a NW of ~74 nm in diameter and a source-drain spacing of $L \sim 120$ nm. Signals along $V_{ds} = 0$ drop to the noise floor for $V_{bg} <$ 1.28 V, although it is not the case of the last electron occupation on the dot. The addition energy exhibits large variations, from ~3 meV to ~10 meV, which can be attributed to large level quantization in the few-electron regime. In the gate voltage region of 1.34 V $< V_{bg} <$ 1.49 V, the transport is characterized by a distorted Coulomb diamond. In particular, a saw-tooth structure is seen to appear at the edge of the diamond. This saw-tooth structure, which is more clearly seen on the negative bias voltage side, could most likely be the effect of the population of remote, localized charge traps surrounding the QD, e.g., in the gate dielectrics. Horizontal features seen clearly inside this distorted Coulomb diamond are signs of co-tunneling involving an excited state.

In order to extract the spin properties of quantum levels in the MBE-grown NW QD, we show, in Figs. 3b and 3c, the evolution of the differential conductance $dI_{ds}/dV_{ds}$ of the device with increasing magnetic field at two fixed gate voltages $V_{bg} = 1.57$ V and 1.62 V, i.e., along the white dashed line cuts A and B in Fig. 3a. At $V_{bg} = 1.57$ V, the QD is occupied by an even number of electrons in the CB region and the pair of inner $dI_{ds}/dV_{ds}$ peaks, sitting symmetrically around $V_{ds} = 0$ as seen in Figure 3b, flags the transport through the two last occupied spin-degenerate orbital levels. With increasing $B$, the two peaks start to move towards each other, reflecting that the spin degeneracy is lifted through the Zeeman effect and the spin-up state (taking the fact that the g-factor is negative into account) moves up with increasing $B$. Here it is also clearly seen that a new peak is split out on the positive $V_{ds}$ side and shifts to higher $V_{ds}$ with increasing $B$. This split peak is associated with the transport through the Zeeman split spin-down state in the QD. On the contrary, at $V_{bg} = 1.62$ V (Figure 3c), the QD is occupied by an odd number of electrons in the CB region and the pair of the inner $dI_{ds}/dV_{ds}$ peaks resemble the transport through the last occupied spin-up orbital level. When the magnetic field is applied, the two inner peaks move apart with increasing $B$. In this case, the two peaks do not exhibit splitting, confirming the half filling of an orbital state with a spin degeneracy of 2. By fitting the peak shift in the low field region to the spin-1/2 Zeeman energy term $E_Z = \pm\frac{1}{2}|g^*|\mu_B B$, where $\mu_B$ is the Bohr magneton and $g^*$ is the effective g-factor, we can extract the g-factors associated with the two neighboring quantum levels as $|g_A^*| = 18$ and $|g_B^*| = 42$.

We can also determine the electron g-factors for different quantum levels from the magnetospectroscopy of the ground states of the quantum dot.[8,33] Here we use the lock-in technique with a small excitation voltage of 5 μV to measure the differential conductance $dI_{ds}/dV_{ds}$ as a function of $V_{bg}$ and $B$ in a more positive gate voltage region containing three consecutive pairs of small and large CB diamonds or, equivalently, involving three neighboring quantum levels. Figure 4 shows the results of the measurements where the evolutions of six consecutive Coulomb peaks with increasing $B$ are presented. As the field is increased, the conductance peaks move up or down, depending on the spin of the last occupied electron on the dot. Since the total ground-state spin state alternates between a singlet $S = 0$ and a doublet $S = 1/2$ in the zero and low magnetic field region, the peak spacing between a pair of a spin-up and a spin-down last occupied electron, after being converted from gate voltage to energy variation via the gate arm factor, reflects the Zeeman energy difference $\Delta E_Z$. By fitting the measured data in the low field region to the Zeeman energy difference $\Delta E_Z = |g^*|\mu_B B$, we can obtain the effective g-factors for the three involved quantum levels. The results are $|g_1^*| = 24$, $|g_2^*| = 30$ and $|g_3^*| = 25$. It is worth noting that in the large $B$ region shown in Figure 4, the movements of the conductance peaks deviate from the linear magnetic field dependence. This is due to an increased influence of the magnetic field on the orbital states and the level repulsion in the presence of spin-orbit interaction (see below) at the crossover in ground states from a singlet to a triplet.[34,35]

Together with the values extracted from the energy spectra shown in Figure 3, the g-factors in our MBE-grown InSb NW QDs show energy level-dependent fluctuations, which is consistent with previous reports on QD devices made from MOVPE-grown InSb NWs[8] and MBE-grown InAs NWs.[36] It has been known that level-to-level fluctuations of the g-factor in a QD imply the presence of spin-orbit coupling in the QD.[37-39] We quantify the spin-orbit coupling strength in our InSb NW QDs by investigating the evolution of the excited state spectroscopy with increasing magnet field. The spin-orbit coupling mixes the Zeeman-split spin-up and spin-down states of different orbitals at a point of degeneracy and gives rise to level anticrossing.[18,40,41] Figure 5a shows the stability diagram of an MBE-grown InSb NW QD device with ~50 nm in the NW diameter and ~220 nm in the contact spacing. The CB diamonds are regularly spaced and show clear even-odd electron-number dependent alternations in size. From the measurements, we extract for the QD device an average charging energy $E_C = 4.5$ meV and a quantum level spacing $\Delta\varepsilon = 1.6$ meV. Figure 5b shows the differential conductance spectrum measured along cut A in the stability diagram shown in Figure 5a as a function of the magnetic field. The spin states of the two involved quantum levels (one in the ground state and the other one in the first excited state) are labeled in Figure 5b. When the magnetic field is increased, the spin-up and spin-down levels first move towards each other, then undergo an avoided crossing at $B \sim 0.4$ T, and finally move apart. It can be seen that the absolute values of the slopes corresponding the two levels are different, but remain almost unchanged after going through the avoided crossing region. The anti-crossing manifest an admixture of spin states near the degenerate point as a result of spin-orbit interaction. By fitting the energy positions of the two peaks to a simple two-level perturbation model, we can extract the spin-orbit coupling strength $\Delta_{SO}$. The data points together with the best fit curves are displayed in Figure 5c. Based on the fitting, we deduce a spin-orbit coupling strength $\Delta_{SO} \approx 275$ μeV and a g-factor $|g^*| = 27$ for the quantum level in the ground state and a g-factor $|g^*| = 48$ for the quantum level in the excited state.

Finally, we would like to note that, considering the very high vacuum chamber and pure elemental sources used in the MBE growth, the coherent nature of electron transport through single InSb NW QDs demonstrated in the present work makes MBE-grown InSb nanowires attractive for further studies of many complex quantum transport phenomena. Especially, in the recent pursuit of Majorana bound states in semiconductor NWs, theoretical studies have proposed alternative interpretations for the experimental observations of zero-bias conductance peaks in superconductor-semiconductor NW hybrid devices and shown that a zero-bias conductance peak can occur in such hybrid devices with a moderate amount of disorder in the semiconductor NWs.[42,43] In this regard, a clean and coherent semiconductor system, such as an MBE-grown InSb NW, is desirable for the experimental detection of Majorana fermions in solid state. Furthermore, the realization of a few-electron quantum system using such a clean and low-disorder InSb NW is important for exploring correlated electron phenomena at a low carrier density, such as the formation of a Wigner crystal at a one-dimensional quantum system,[44-46] where the presence of only a small, finite number of electrons is crucial.

In conclusion, we have performed a detailed spectroscopic study of single QDs of different sizes made from MBE-grown InSb nanowires and have demonstrated coherent single-electron transport in the devices. The nanowires have a diameter of 50-80 nm and the QDs are defined in the nanowires between contacts by naturally formed Schottky barriers. Our results show that high-quality single QDs with a longitudinal size up to 700 nm can be realized using the MBE-grown InSb nanowires and the electron transport in the QDs can be controlled to occur in both the many-electron and the few-electron regime. The linear relationship between the gate capacitance and the source-drain electrode separation provides an experimental evidence for electron tunneling through quantum levels in single QDs. The measured g-factors are extracted to be in a range of 18-48 and show level-dependent fluctuations. A spin-orbit coupling strength of $\Delta_{SO} \sim 275$ μeV is deduced by measuring the anticrossing of a ground state and a neighboring excited state. These results establish MBE-grown InSb NWs as a highly desired material platform for the further studies of strong-correlated many-body physics, including Majorana fermions in quantum nanostructures and for the development of high-performance quantum information devices.

**Acknowledgement**

This work was supported by the National Basic Research Program of the Ministry of Science and Technology of China (Grants No. 2012CB932703 and No. 2012CB932700), the National Natural Science Foundation of China (Grants No. 11374019, No. 91221202, No. 91421303 and No. 61321001). N.K. thanks the Ph.D. Program Foundation of the Ministry of Education of China for financial support (Grant No. 20120001120126). H.Q.X acknowledges also the financial support from the Swedish Research Council (VR).

**Correspondences**

* Corresponding author: nkang@pku.edu.cn

* Corresponding author: hqxu@pku.edu.cn

**Present address**

† Department of Electronic Materials Engineering, Research School of Physics and Engineering, the Australian National University, Canberra, ACT 0200, Australia

**Notes**

The authors declare no competing financial interests.

**98**, 266801.

**Figures and captions:**

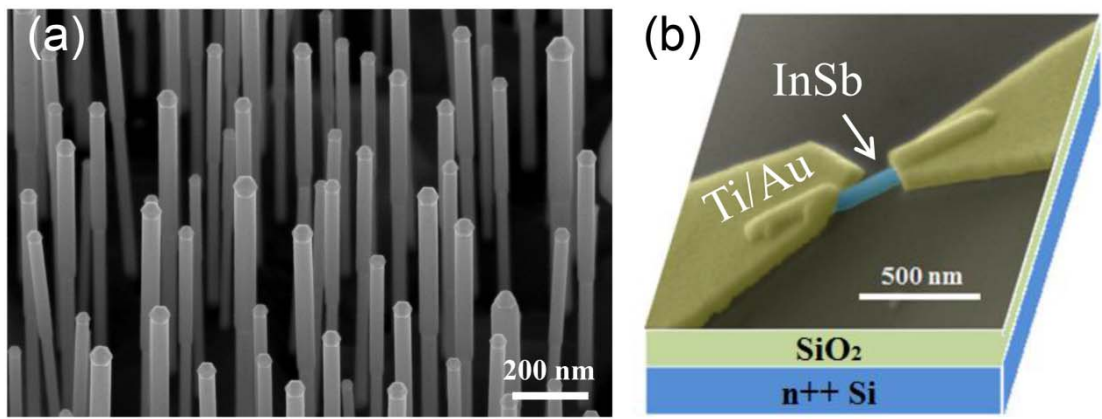

**Figure 1.** (a) Representative SEM image (30° tilted angle view) of the as-grown nanowire sample, illustrating from top to bottom the $AuIn_2$ seed particles, the top InSb segments, and the thinner InAs stems, respectively. (b) SEM image (in false colors) of a typical InSb nanowire device contacted by Ti/Au and schematic view of the $Si/SiO_2$ substrate on which the device is fabricated. An InSb nanowire quantum dot is formed between the two Ti/Au contacts at low temperatures.

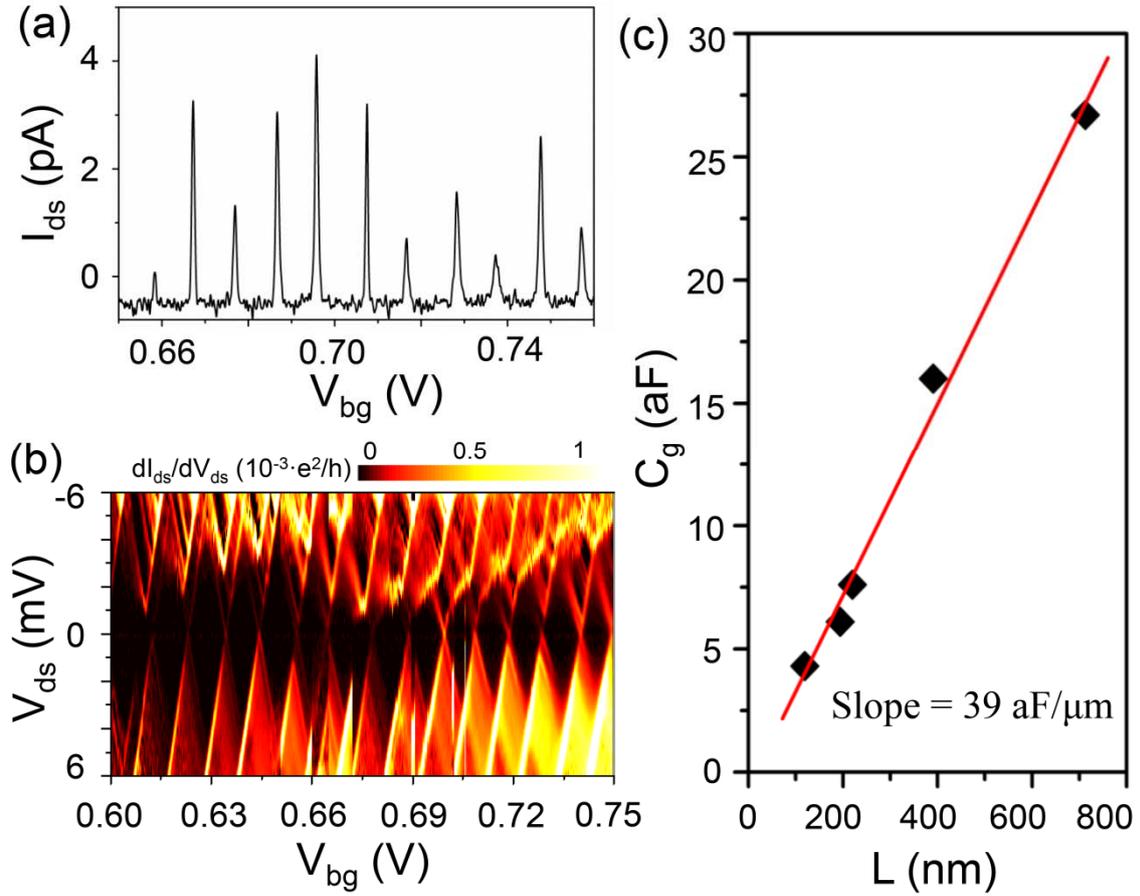

**Figure 2.** (a) Coulomb blockade oscillations of the source-drain current measured for an InSb nanowire quantum dot device with a spacing of 391 nm between the source and drain contacts at applied bias voltage $V_{ds}$ = 0.1 mV. (b) Plot of the differential conductance as a function of back gate voltage $V_{bg}$ and bias voltage $V_{ds}$ (charge stability diagram) for the InSb nanowire quantum dot in the many-electron regime. (c) Gate capacitance extracted as a function of the source-drain spacing $L$. The red line with a slope of 39 aF/μm is a linear fit to the data points extracted from the measurements of five InSb nanowire quantum dot devices.

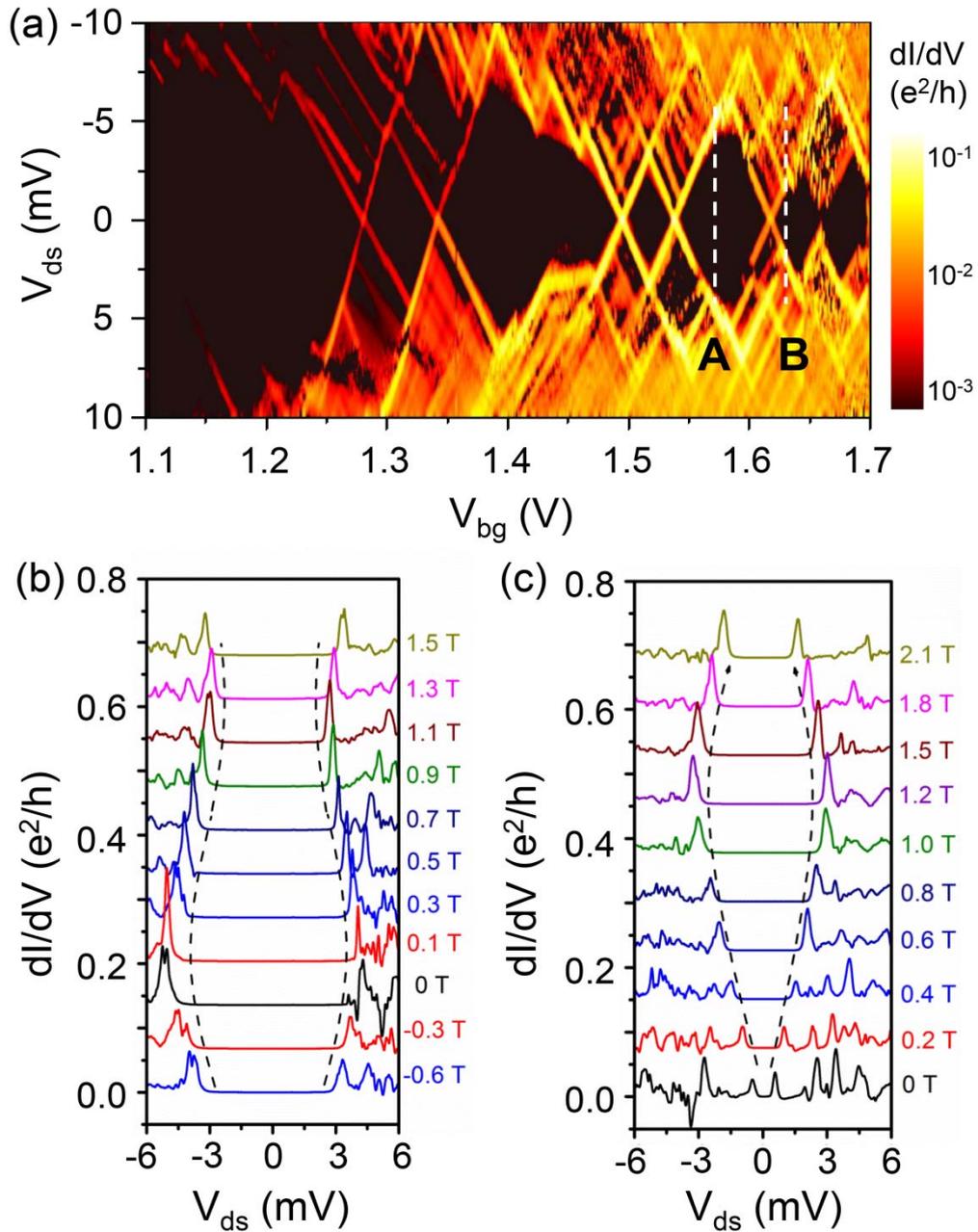

**Figure 3.** (a) Charge stability diagram of an InSb nanowire quantum dot device with a source-drain spacing of ~120 nm in the few-electron regime. (b) Magnetic-field evolution of the differential conductance $dI_{ds}/dV_{ds}$ measured for the device at back gate voltage $V_{bg}$ = 1.57 V, i.e., along the white-dashed line cut A in (a). (c) Magnetic-field evolution of the differential conductance $dI_{ds}/dV_{ds}$ measured for the device at back gate voltage $V_{bg}$ =1.62 V, i.e., along the white-dashed line cut B in (a). The dashed lines in (b) and (c) are guides to the eyes.

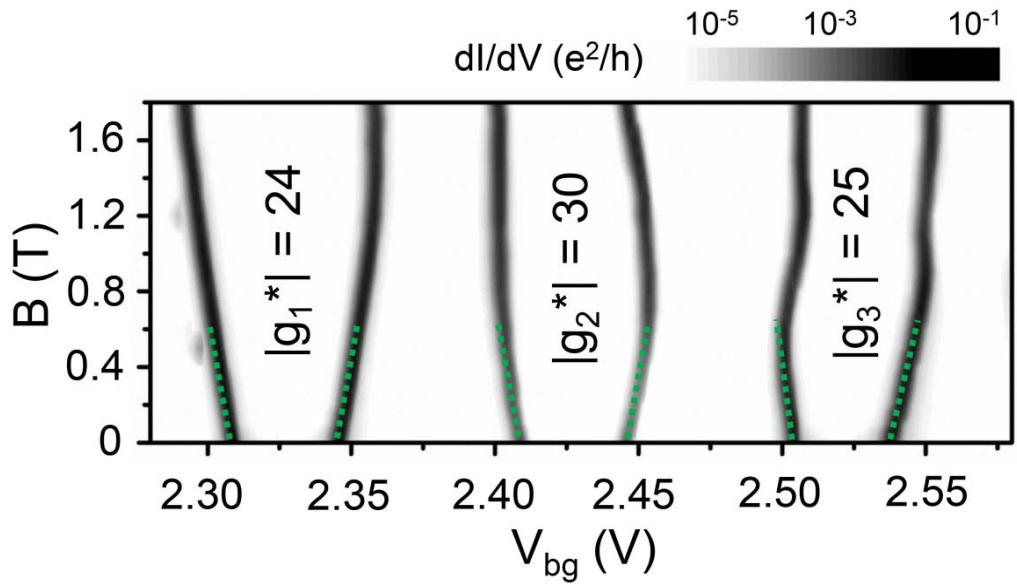

**Figure 4.** Gray scale plot of magnetic field evolution of the differential conductance $dI_{ds}/dV_{ds}$ measured in the linear response regime (i.e., at bias voltage $V_{ds} = 0$ V) for the same quantum dot device as in Fig. 3. The g factors extracted from the measurements for quantum levels are indicated in the figure. The green dashed lines denote the data used to extract the g-factors.

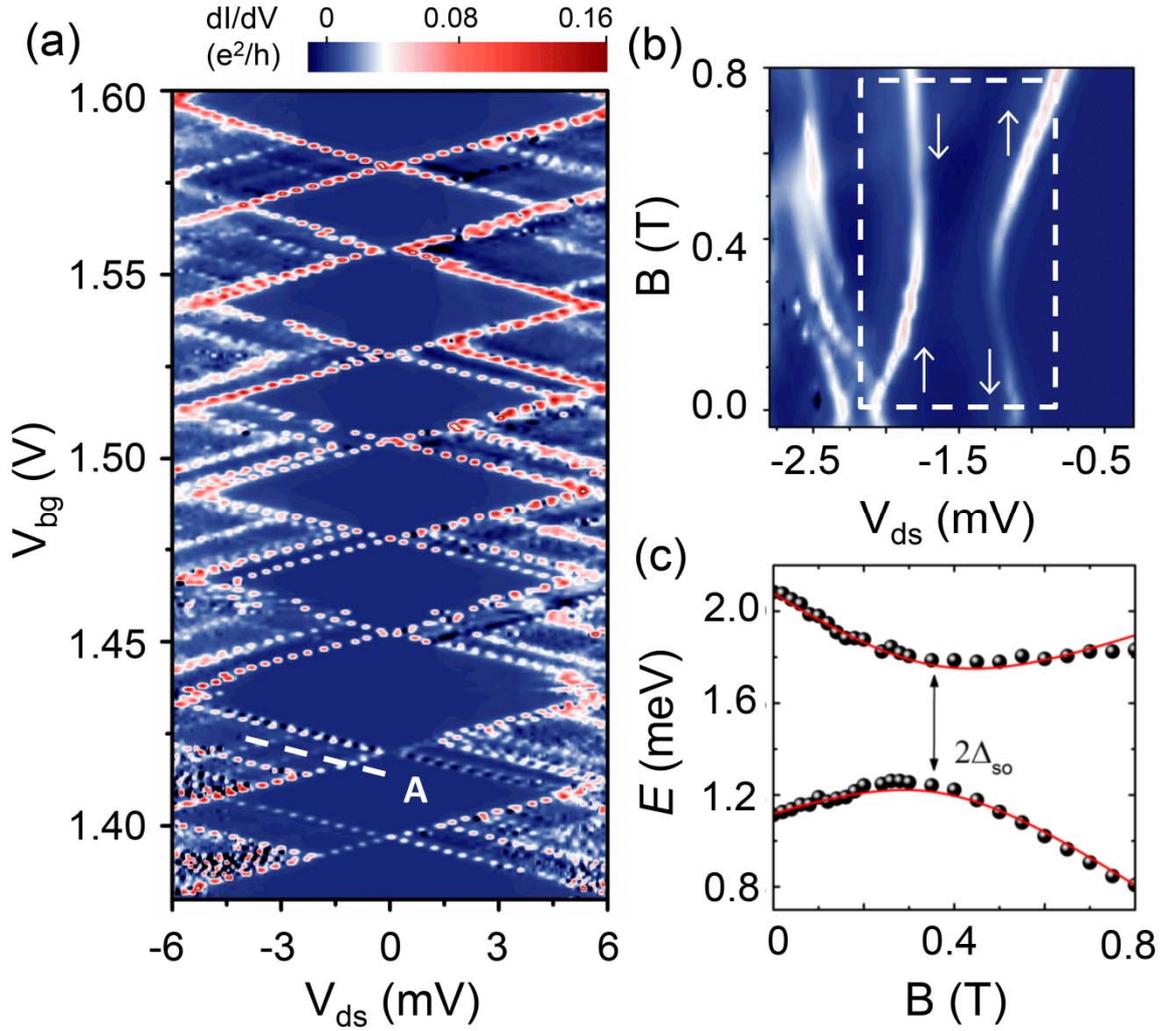

**Figure 5.** Magnetic field evolutions of a ground state and a neighboring excited state of an InSb nanowire quantum dot device with a source-drain contact spacing of ~220 nm. (a) Charge stability diagram of the quantum dot at $B = 0$ T. (b) Differential conductance measured along line cut A in (a) as a function of $B$. (c) Converted energy positions of the differential conductance peaks in the region marked by the dashed rectangle in (b). The dots are the experimental energy positions of the peaks, the red lines are the results of a fit to a two-level perturbation model, and $\Delta_{so}$ is the spin-orbit coupling strength assumed in the model, which causes the two quantum levels with different spins to undergo anti-crossing.